\documentclass[10pt, conference, letterpaper]{IEEEtran}

\IEEEoverridecommandlockouts

\usepackage{cite}
\usepackage{amsmath,amssymb,amsfonts}
\usepackage{algorithm}
\usepackage{float}
\usepackage{graphicx}
\usepackage{quantikz}
\usepackage{textcomp}
\usepackage{subcaption} 
\usepackage{booktabs}
\usepackage{multirow}
\usepackage{siunitx}
\usepackage{pgfplots}
\usepgfplotslibrary{groupplots}
\usetikzlibrary{calc}
\pgfplotsset{compat=1.18}
\usepackage{algpseudocode}
\usepackage[margin=0.5in]{geometry} 
\usepgfplotslibrary{colormaps}

\definecolor{cBlue}{RGB}{31, 119, 180}
\definecolor{cOrange}{RGB}{255, 127, 14}
\definecolor{cRed}{RGB}{214, 39, 40}

\definecolor{cBlue}{RGB}{0, 114, 178}   
\definecolor{cOrange}{RGB}{213, 94, 0}  

\pgfplotscreateplotcyclelist{mycustomlist}{
    {cBlue, very thick},
    {cOrange, very thick},
    {cRed, very thick}
}
\usepackage{xcolor}
\def\BibTeX{{\rm B\kern-.05em{\sc i\kern-.025em b}\kern-.08em
    T\kern-.1667em\lower.7ex\hbox{E}\kern-.125emX}}

\begin{document}

\title{Generating probability distributions using variational quantum circuits
}

\author{\IEEEauthorblockN{Ronit Raj\IEEEauthorrefmark{1},
Kshitij Durge\IEEEauthorrefmark{2},
Manish Mallapur\IEEEauthorrefmark{3},
Rohit Taeja Kumar\IEEEauthorrefmark{4} and
Ankur Raina\IEEEauthorrefmark{5}}
\IEEEauthorblockA{Dept. of Electrical Engineering and Computer Science, Indian Institute of Science Education and Research Bhopal, India\\
Email: \IEEEauthorrefmark{1}ronit21@iiserb.ac.in,
\IEEEauthorrefmark{2}durge21@iiserb.ac.in,
\IEEEauthorrefmark{3}manish21@iiserb.ac.in,
\IEEEauthorrefmark{4}r.taeja@gmail.com,
\IEEEauthorrefmark{5}ankur@iiserb.ac.in}}

\maketitle

\begin{abstract}
Sampling from a probability distribution is a core task in many quantum and classical algorithms. Variational quantum circuits provide a natural approach to generating such distributions, as measurement outcomes directly define the probability values.
However, designing circuits that train reliably while utilizing limited quantum resources remains largely a heuristic approach. In particular, the roles of expressibility, entanglement capability, and quantum resources in training performance and scalability are not well understood. 
In this work we present a systematic study of variational quantum circuits where we compare different ansatze family across multiple cost functions and classical optimization methods. We use expressibility and entanglement capability as circuit descriptors to explain convergence behaviors, optimizer sensitivity and robustness to noise. Our results provide a practical guidelines for designing resource aware, efficient and trainable quantum circuits, moving beyond heuristic methods for near term applications.
\end{abstract}


\section{Introduction}

The ability to generate an arbitrary probability distribution is a foundational task that has numerous applications in quantum computing, from financial modeling \cite{wilkens2023quantum} to state preparation in quantum machine learning and Hamiltonian simulation. 
Given that quantum circuits are inherently probabilistic, they present a natural and powerful framework for this task \cite{grover2002creating, dasgupta2022loading}. 
In the noisy intermediate-scale quantum (NISQ) era \cite{preskill2018quantum}, hybrid-quantum algorithms\cite{cerezo2021variational} emerge as the leading algorithmic framework for practical quantum computing applications. 
The success of these algorithms is critically dependent on the expressiveness and trainability of Parametrized quantum circuits (PQC)\cite{hubregtsen2021evaluation}.
However, despite their potential to generate a probability distribution, there is a lack of a systematic understanding of the principles that connect a PQC's architecture and training techniques to its generative performance. 
Understanding the connection between expressibility and the entanglement properties\cite{sim2019expressibility} of a PQC becomes a key factor that affects its trainability. Classical optimization also affects the accuracy and trainability of a PQC.

In this work, we address this challenge by providing a systematic benchmarking study of variational quantum circuits in probability distribution generation.
We explicitly characterize different families of PQC ansatz based on their expressibility, entanglement capability, and resource requirements.
By comparing these families of ansatze across multiple cost functions and classical optimizers, we establish clear connections between circuit properties, trainability, and final convergence.
This enables a more principled understanding of why some ansatz succeed or fail under specific tasks. 

Our approach combines variational training experiments with scaling and noise analysis for near-term feasibility.
We evaluate performance across various qubit counts and noise models, highlighting trade-offs between accuracy, trainability, and resource requirements.
These benchmarks provide guidance for informed ansatze selection.

The rest of the paper is organized as follows. Section~\ref{sec:background} reviews key metrics. Section~\ref{sec:method} describes the ansatz families, training protocol, and noise protocol. Section~\ref{sec:results} presents the benchmarks and analysis. We conclude in Section~\ref{sec:concl}.


\section{Background}
\label{sec:background}

Variational Quantum Algorithms (VQAs) leverage the variational principle of quantum mechanics to solve optimization problems by iteratively optimizing a cost function $C(\theta)$ through a hybrid quantum-classical framework.
This process employs an ansatz $U(\theta)$, to prepare a trial state, whose properties are measured to evaluate the cost function. 
A classical optimizer then updates the parameters $\theta$ to traverse the optimization landscape defined by the cost function. 
The performance of VQA depends on two major components: the cost function, which must be faithful, feasible, and sufficiently smooth for optimization, and the ansatz architecture.
The ansatz design presents a fundamental tradeoff between expressibility, a circuit's ability to span the Hilbert space; and trainability, the practical feasibility of optimizing its parameters. 
While high expressibility is a prerequisite for high accuracy, it often hinders the performance with barren plateaus, where cost function gradients vanish exponentially with system size, impacting the performance of classical optimizers. 
Consequently, the central challenge is the expressibility-trainability balance. Similarly, the choice of cost functions and classical optimizers also dictates the performance of VQA.

\subsection{Expressibility of a parametrized quantum circuit}

Expressibility characterizes a PQC's ability to generate diverse quantum states. In this work, we adopt the fidelity-based method introduced by \cite{sim2019expressibility}, as it is an operationally meaningful and practically estimable measure of how well a circuit can explore the Hilbert space. A circuit is said to be highly expressible if it generates states that can approximate the uniform Haar distribution. We quantify this by comparing the fidelity distribution ($F$) of the PQC generated states, $\hat{P}_{\text{PQC}}(F)$, against the theoretical distribution of Haar random states, $P_{H}(F)$
\begin{equation}
    F = |\langle \psi(\theta) | \psi(\phi) \rangle|^2.
\end{equation}

Repeating this procedure many times yields an empirical distribution $\hat{P}_{\text{PQC}}(F)$ of fidelities. This is compared against the known fidelity distribution of Haar-random states,
\begin{equation}
    P_H(F) = (N-1)(1-F)^{N-2},
\end{equation}

\noindent where $N = 2^n$ is the Hilbert space dimension. 
Sim \emph{et al.} originally proposed using the Kullback–Leibler (KL) divergence between $\hat{P}_{\text{PQC}}(F)$ and $P_\text{Haar}$ distributions to quantify expressibility.
In our work, we use the Jensen-Shannon divergence ($D_{JSD}$), a symmetric and bounded alternative to the KL divergence.

        \begin{equation}
        E = -\log_{10}(D_{\text{JSD}}(\hat{P}_{\text{PQC}}(F) \parallel P_{\text{Haar}}(F))),\label{eq:exp}
       \end{equation}

where $E$ is the negative base-10 logarithm of divergence to create a more intuitive metric where higher values correspond to greater expressibility.

\subsection{Entangling Capability of Parameterized Quantum Circuits}

A key descriptor for PQCs is entangling capability, defined as a circuit's ability to generate entangled states. 
This is crucial for capturing complex correlations present in the data. 
To quantify this, we employ the Meyer-Wallach (MW) measure as described in \cite{sim2019expressibility}, $\mathcal{Q}$, as it is computationally efficient and scalable. 
For a given $n$-qubit quantum state $|\psi\rangle$, it is defined as: 
\begin{align}
    \mathcal{Q}(|\psi\rangle) \equiv \frac{4}{n} \sum_{j=1}^{n} D(\iota_{j}(0)|\psi\rangle, \iota_{j}(1)|\psi\rangle), 
\end{align} 
that assigns a score of $0$ to completely unentangled states and 1 to maximally entangled states. To evaluate the a PQC's overall capability, we estimate the average entanglement it can produce. The average entanglement capability, $\mathcal{E}$, is then quantified by averaging $\mathcal{Q}$ over a representative set $S$ of circuit parameters $\theta$:
 \begin{align}
     \mathcal{E} = \frac{1}{|S|} \sum_{\theta \in S} \mathcal{Q}(|\psi(\theta)\rangle).
 \end{align}
This results in a single score from 0 to 1, providing a robust metric to compare the entangling power of different PQC architectures.


\section{Methodology}
\label{sec:method}

Our methodology is designed to evaluate how different parameterized quantum circuit (PQC) designs perform for the task of generating probability distributions. Our main emphasis is on circuit descriptors like expressibility and entanglement capabliity effect trainability, robustness, and scaling.
We compare three ansatz families and benchmark them across several loss functions and classical optimizers. 
Below we summarize the ansatz construction and characterization, the variational training protocol, and our noise and scaling experiments.

\subsection{Ansatz families and characterization}
\label{subsec:ansatz}
We evaluate three distinct ansatz families:
\begin{enumerate}
  \item Predefined structured ansätze: These circuits are taken from prior literature \cite{sim2019expressibility}, composed of fixed parameterized single-qubit layers and predefined entangling patterns repeated to form the full architecture.
  \item Random circuits: Architectures where each layer consists of independently and uniformly sampled single- and two-qubit gates, producing unstructured parameterizations, with a final predefined entangling block \cite{wu2024randomness}.

  \item Enhanced random circuits: Circuits adapted from prior work in which random architectures are evolved using a genetic framework \cite{mallapur2025genetic} to maximize expressibility, followed by the inclusion of a fixed entangling layer.

\end{enumerate}

\begin{figure}[htbp]
  \centering
  \includegraphics[width=\columnwidth]{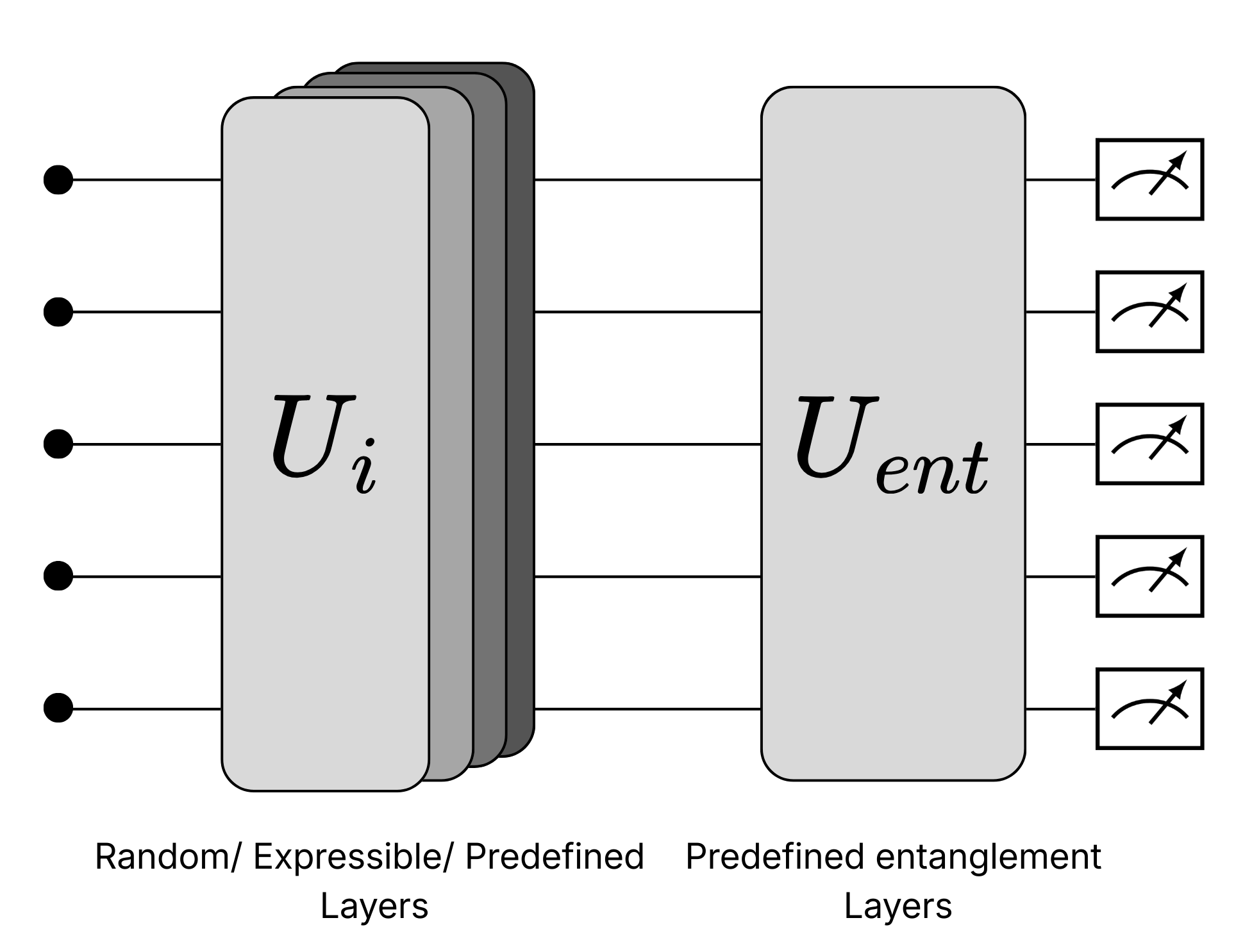}
  \caption{Circuit template used throughout this work: stacked parameterized single-qubit layers (random, expressible, or predefined) followed by a fixed entangling block $U_{\mathrm{ent}}$. This template is used to construct all three ansatz families.
}
  \label{fig:circuit}
\end{figure}

Each instance is represented as $U(\theta)$ formed by stacked single-qubit parameterized layers $U_i$ followed by an entangling block $U_{\mathrm{ent}}$ (see Fig.~\ref{fig:circuit}). For each circuit we record: (i) circuit depth, (ii) two-qubit gate count, (iii) expressibility, and (iv) entanglement capability. These descriptors are used to interpret observed training behaviour.

\subsection{Variational training protocol}
\label{subsec:training}
The objective is to minimize a divergence-based loss $C\big(P_{\theta},P_{t}\big)$ between the PQC output distribution $P_{\theta}$ and a target distribution $P_{t}$. We evaluate the following loss functions: KL-divergence, Jensen-Shannon divergence (JSD), and Hellinger distance. For optimization we compare gradient-based methods (Adam, L-BFGS-B) and gradient-free methods (CMA-ES, Metropolis).

\begin{algorithm}[H]
\caption{Variational Quantum Distribution Generation}
\label{alg:vqdg_short}
\begin{algorithmic}[1]
\State \textbf{Input:} Target distribution $P_t$
\State \textbf{Hyperparameters:} Optimizer settings, number of steps $S$
\State \textbf{Output:} Optimized parameters $\boldsymbol{\theta}^*$

\State Initialize parameters $\boldsymbol{\theta}$ randomly
\For{$s = 1$ to $S$}
    \State Execute PQC $U(\boldsymbol{\theta})$ to obtain $P_{\boldsymbol{\theta}}$
    \State Compute cost $C(P_{\boldsymbol{\theta}}, P_t)$
    \State Update $\boldsymbol{\theta}$ using a classical optimizer
\EndFor
\State \textbf{return} $\boldsymbol{\theta}^*$
\end{algorithmic}
\end{algorithm}

\begin{figure}[htbp]
  \centering
    \includegraphics[width=\columnwidth]{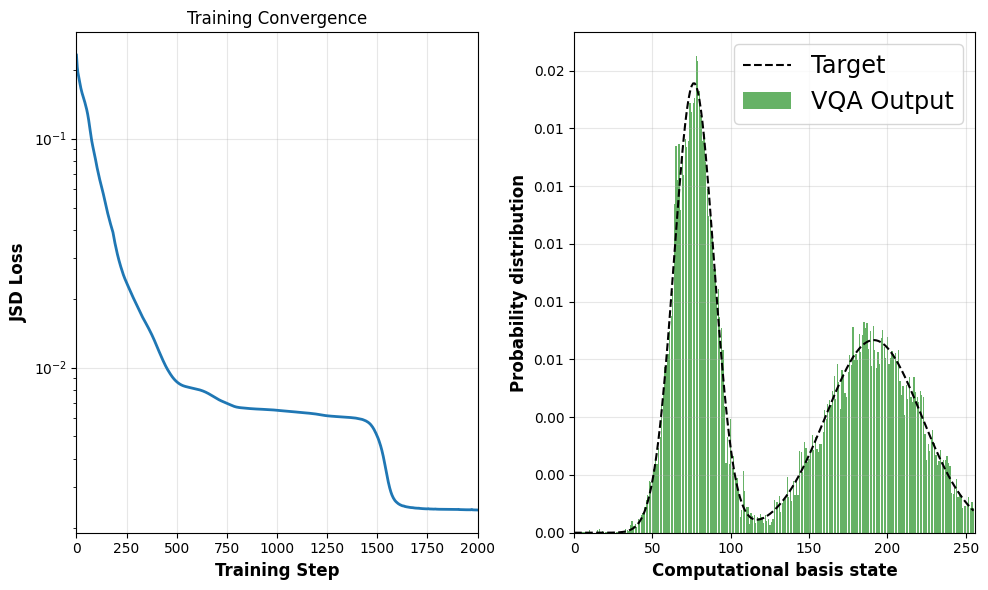}

  \caption{Left: Training convergence showing Jensen–Shannon divergence (JSD) loss versus optimization step for an 8-qubit high-expressibility predefined ansatz trained with the Adam optimizer on an asymmetric bimodal target distribution. Right: final variational quantum ansatz (VQA) output distribution (bars) compared to the target distribution (dashed).
}
  \label{fig:training}
\end{figure}

Figure \ref{fig:training} shows an example of the variational training protocol, where an 8-qubit predefined ansatz is trained to generate an asymmetric bimodal probability distribution.

\subsection{Noise}
\label{subsec:noise}
To assess near-term viability, a subset of experiments was repeated under simple noise channels: per-gate depolarizing noise, amplitude damping, phase damping, and readout bit-flip errors. Noise parameters were chosen to reflect existing noisy device behaviour, like the relaxation and dephasing times and readout error rates. Experiments were performed on 4 qubits with twenty initialisation seeds, and results were averaged across the seeds to evaluate performance under noisy conditions.

This setup allows us to compare different ansatz families in a consistent way by relating their training performance to circuit properties. In the next section, we present results that demonstrate how expressibility and entanglement impact convergence, optimizer behavior, and scaling across varying problem sizes.


\section{Results}
\label{sec:results}

We organize the results in three major points of interest :  (i) How circuit level descriptors (expressibility and entanglement capability) affect training dynamics and final fidelity, (ii) How optimizer and cost choices relates with those circuit descriptors, (iii) How quantum resource usage like circuit depth and gate count scales across ansatz families, and how these resource requirements affect their feasibility under noisy simulations. 
We report training performance of these three ansatz families on various distribution (Uniform, Normal, Bimodal Gmm and Arbitrary), cost function (Kullback Leibler divergence, Jensen Shanon divergence and Hellinger distance ) and optimizers (CMA-ES, Metropolis, Adam, and L-BFGS-B).

\subsection{Ansatz Performance Comparison}
\label{subsec: APC}

We begin by comparing the performance across the three ansatze families-Predefined, Random and Enhanced circuits on generating probability distributions. All circuits are evaluated on the same distribution using same cost function and optimizers to ensure fair comparison. 

Across both 4 qubit and 8 qubit systems, random circuits performed the weakest. The training loss function stagnates with JSD divergence values around values $10^-{1}$ or higher. This behavior is observed across all distributions indicating limited representational power and poor trainability. Whereas predefined circuits with varied expressibility and entanglement show mixed results. Circuits with low expressibility and high entanglement struggle to converge well for distribution like normal, bimodal and arbitrary. In contrast circuits with high expressibility are able to converge to substantially lower final divergence and show stable convergence across all tested targets.

Expressibility enhanced circuits were able to perform more consistently over all systems. These circuits converged to final divergence values of order of approximately $10^{-3}$ or lower using JSD cost function and metropolis as the classical optimizer. Figure 3 shows that this improvement is achieved only when high expressibility is accompanied by moderate entanglement capability. Circuits with very high entanglement and limited expressibility do not show similar gains while circuits with high expressibility and moderate entanglement converge reliably across all distributions.

These results show that expressibility is a key factor in achieving high accuracy but only when supported with sufficient entanglement. This balance enables the circuits to represent complex distribution


\begin{figure}
    \centering
    \includegraphics[width=\columnwidth]{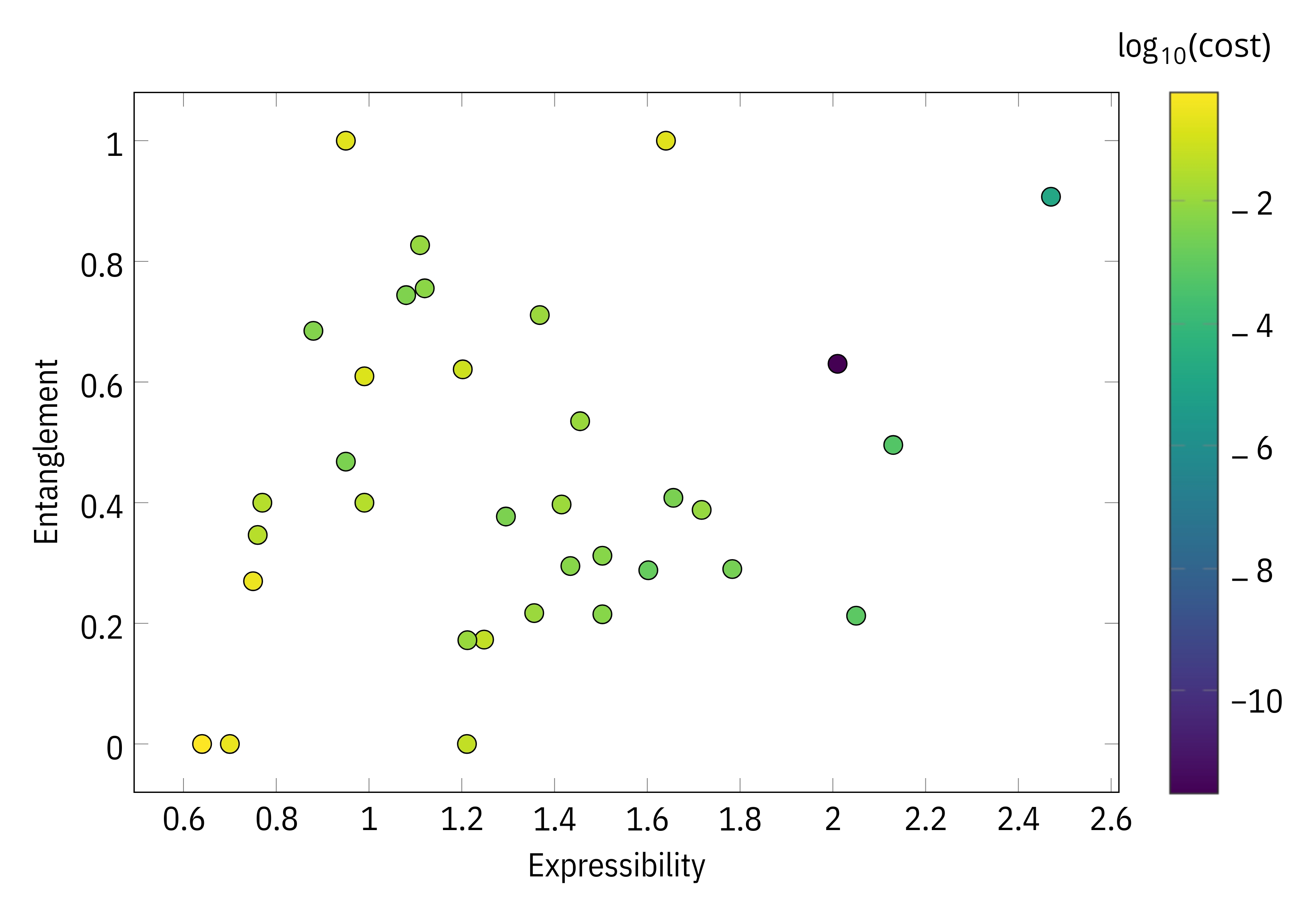}
    \caption{ Relationship between expressibility, entanglement capability, and final training performance across all evaluated circuits. Each point corresponds to a circuit instance from the predefined, random, or expressibility-enhanced families. High accuracy is achieved primarily by circuits combining high expressibility with moderate entanglement.
}
    \label{fig:placeholder}
\end{figure}

\begin{figure*}[ht]
    \centering
    \begin{tikzpicture}
        \begin{groupplot}[
            group style={
                group size=3 by 1,
                horizontal sep=0.5cm,
                y descriptions at=edge left,
                xlabels at=edge bottom
            },
            width=0.32\textwidth, height=5cm,
            %
            xmode=log, ymode=log,
            grid=major,
            grid style={dashed, gray!30},
            xmin=1, xmax=100000,
            ymin=1e-5, ymax=1.5,  
            xlabel={\textbf{Iterations}},
            tick label style={font=\footnotesize},
            label style={font=\small\bfseries},
            %
            cycle list name=mycustomlist,
            %
            legend columns=3,
            legend style={
                draw=none, 
                fill=none, 
                font=\small,
                /tikz/every even column/.append style={column sep=0.5cm}
            }
        ]

        \nextgroupplot[
            title={\textbf{Predefined}},
            ylabel={\textbf{Cost (JSD)}}
        ]
        \addplot table [x=steps, y=Circuit1, col sep=comma] {QCNCImages/data_Predefined.csv};
        \addplot table [x=steps, y=Circuit2, col sep=comma] {QCNCImages/data_Predefined.csv};
        \addplot table [x=steps, y=Circuit3, col sep=comma] {QCNCImages/data_Predefined.csv};

        \nextgroupplot[
            title={\textbf{Random}},
            legend style={at={(0.5,-0.30)}, anchor=north}
        ]
        \addplot table [x=steps, y=Circuit1, col sep=comma] {QCNCImages/data_Random.csv};
        \addlegendentry{Circuit 1}
        
        \addplot table [x=steps, y=Circuit2, col sep=comma] {QCNCImages/data_Random.csv};
        \addlegendentry{Circuit 2}
        
        \addplot table [x=steps, y=Circuit3, col sep=comma] {QCNCImages/data_Random.csv};
        \addlegendentry{Circuit 3}

        \nextgroupplot[title={\textbf{Enhanced}}]
        \addplot table [x=steps, y=Circuit1, col sep=comma] {QCNCImages/data_Enhanced.csv};
        \addplot table [x=steps, y=Circuit2, col sep=comma] {QCNCImages/data_Enhanced.csv};
        \addplot table [x=steps, y=Circuit3, col sep=comma] {QCNCImages/data_Enhanced.csv};

        \end{groupplot}
    \end{tikzpicture}
    \caption{Convergence analysis using metropolis optimizer and JSD metric. The plots compare predefined, random, and expressibility enhanced 4-qubit architectures' performance over the normal distribution. The metropolis optimizer ran for 50000 iterations. Enhanced architectures achieve lower orders of magnitude outperforming random architectures.}
    \label{fig:convergence_comparison}
\end{figure*}
\subsection{Resource Estimation \& Scaling}

We now examine how the performance trends observed in \ref{subsec: APC} correlates with qauntum resource usage such circuit depth and gate counts. These quantities are a key indicator of near term feasibility. Table~\ref{tab:resource_table} summarizes the resource requirements and final convergence values for 4-qubit and 8-qubit systems across all three ansatze families. Random circuits employed in the study have shallow depth and few two qubits gates, this resource shortage does not translate well in performance. The resource shortage reduces the representational power of the circuit to generate complex distributions. Predefined circuits exhibit a wide range of behaviors depending on their architecture. Circuits with high expressibility tend to converge to low final cost values but at the expense of very high circuit depth. Expressibility enhanced circuits provide a favourable middle ground. As shown in the table~\ref{tab:resource_table} these circuits achieve comparable performance to highly expressible predefined circuit while using  fewer two qubit gates and much shallower depths.  This highlights a key tradeoff: reducing resources too aggressively limits the representational power, while excessive depth results in diminishing returns but also makes it infeasible for near term applications. Scaling from 4 qubit to 8 qubit further amplifies these trends.

Overall, these results demonstrate that effective resource usage can be utilized by aligning to the circuit expressibility and entanglement capability. Expressibility enhanced circuits provide a more favorable resource-accuracy tradeoff, making them suitable for near-term implementations.

\begin{table}[htbp]
    \centering
    \caption{Resource estimation and performance benchmarking for the adam optimizer and jsd metric over the normal distribution. The notations are as follows: $n$: number of qubits, Expr: expressibility, $N_{2q}$: number of two-qubit gates, $d$: circuit depth.}
    \label{tab:resource_table}
    \renewcommand{\arraystretch}{1.3} 
    \setlength{\tabcolsep}{5pt}
    \vspace{-0.1cm}
    \resizebox{\columnwidth}{!}{%
    \begin{tabular}{|l c c c c c c|}
        \hline
        \textbf{Set} & \boldmath$n$ & \textbf{ID} & \textbf{Expr.} & \boldmath$N_{2q}$ & \boldmath$d$ & \textbf{Cost} \\
        \hline
        
        \multirow{6}{*}{Predefined} 
            & \multirow{3}{*}{4} & 1  & 2.01 & 24   & 45  & \num{2.73e-4} \\
            &                    & 2  & 0.95  & 3    & 5   & \num{1.82e-1} \\
            &                    & 3  & 2.05  & 16    & 27  & \num{2.85e-4}   \\
            \cline{2-7} 
            & \multirow{3}{*}{8} & 1 & 2.47  & 224  & 303 & \num{1.25e-2} \\
            &                    & 2 & 1.64  & 14   & 18  & \num{1.75e-1} \\
            &                    & 3 & 2.13  & 64   & 101  & \num{3.73e-03} \\
        \hline
        
        \multirow{6}{*}{Random} 
            & \multirow{3}{*}{4} & 1 & 0.70  & 4 & 5 & \num{2.84e-01} \\
            &                    & 2 & 1.08 & 10 & 9 & \num{3.62e-02} \\
            &                    & 3 & 0.64 & 5 & 7 & \num{5.94e-01} \\
            
            \cline{2-7}
            & \multirow{3}{*}{8} & 1 & 0.75 & 25 & 9 & \num{3.05e-01} \\
            &                    & 2 & 0.77 & 22 & 9 & \num{3.35e-01} \\
            &                    & 3 & 0.88 & 41 & 17& \num{7.06e-02} \\
        \hline
        
        \multirow{6}{*}{Enhanced} 
            & \multirow{3}{*}{4} & 1 & 0.76 & 5 & 5 &  \num{3.42e-01}\\
            &                    & 2 & 0.95 & 9 & 9 & \num{3.07e-02} \\
            &                    & 3 & 1.12 & 7 & 9& \num{6.10e-03} \\
            \cline{2-7}
            & \multirow{3}{*}{8} & 1 & 0.99 & 21 & 9 & \num{1.59e-02} \\
            &                    & 2 & 0.99 & 22 & 9 & \num{3.35e-03} \\
            &                    & 3 & 1.11 & 36 & 17 & \num{3.40e-03} \\
        \hline
    \end{tabular}%
    }
    
    \raggedright
\end{table}


\subsection{Optimizer \& Cost Sensitivity}
\label{subsec:opt-cost}

Figure~\ref{fig:optimizer_scaling_small} and Tables~\ref{tab:resource_table}-\ref{tab:best_combo_performance} show three clear patterns in optimizer behaviour. First, L-BFGS-B typically converges fastest on the smaller (4-qubit) problems but loses its advantage as the problem size (and parameter count) increases. Second, CMA-ES achieves the best median/best costs across many targets, but uses many more function evaluations. This is expected because CMA-ES is a gradient-free, population-based search that is robust to nonconvexity and noisy gradients, albeit at the expense of being more computationally expensive. Third, the Jensen–Shannon divergence produces smoother, more stable loss landscapes than KL or Hellinger, since JSD is symmetric, bounded, and less sensitive to zero-probability bins, those numerical properties reduce extreme gradients and make optimization easier, particularly for gradient-based methods. Therefore, we can conclude that L-BFGS-B works well when the parameter space is small and gradients are reliable, CMA-ES works well when robustness matters and evaluations are cheap, and JSD-like losses are better when numerical stability is a concern.

\begin{figure}[ht]
    \centering
    \begin{tikzpicture}
        \begin{groupplot}[
            group style={
                group size=2 by 1,
                horizontal sep=0.4cm,
                y descriptions at=edge left,
                xlabels at=edge bottom
            },
            width=5.3cm, height=8cm,
            xmode=log, ymode=log,
            grid=major,
            grid style={dotted, gray!50},
            xlabel={\textbf{Iterations}},
            ylabel={\textbf{Cost (JSD)}},
            tick label style={font=\tiny},
            label style={font=\scriptsize\bfseries},
            title style={font=\scriptsize\bfseries, yshift=-0.8ex},
            cycle list={
                {cBlue, thick},           
                {cBlue, thick, dashed},   
                {cOrange, thick},         
                {cOrange, thick, dashed}  
            },
            legend style={
                font=\small,             
                draw=none,              
                fill=none,             
                fill opacity=0.1,       
                text opacity=0.8,         
                inner sep=0pt,          
                outer sep=0pt,          
                row sep=-2pt,           
                nodes={inner sep=1pt}   
            }
        ]

        \nextgroupplot[
            title={Gradient-Based},
            xmin=1, xmax=1000,
            legend pos=south west
        ]
        
        \addplot+[mark=none, each nth point=1] table [x=steps, y=adam_4q, col sep=comma] {QCNCImages/data_grad_based.csv};
        \addlegendentry{Adam-4q}
        \addplot+[mark=none, each nth point=1] table [x=steps, y=adam_8q, col sep=comma] {QCNCImages/data_grad_based.csv};
        \addlegendentry{Adam-8q}
        \addplot+[mark=none, each nth point=1] table [x=steps, y={l-bfgs-b_4q}, col sep=comma] {QCNCImages/data_grad_based.csv};
        \addlegendentry{L-BFGS-4q}
        \addplot+[mark=none, each nth point=1] table [x=steps, y={l-bfgs-b_8q}, col sep=comma] {QCNCImages/data_grad_based.csv};
        \addlegendentry{L-BFGS-8q}

        \nextgroupplot[
            title={Gradient-Free},
            xmin=10, xmax=50000,
            legend pos=south west
        ]
        
        \addplot+[mark=none, each nth point=20] table [x=steps, y=cma_4q, col sep=comma] {QCNCImages/data_grad_free.csv};
        \addlegendentry{CMA-4q}
        \addplot+[mark=none, each nth point=20] table [x=steps, y=cma_8q, col sep=comma] {QCNCImages/data_grad_free.csv};
        \addlegendentry{CMA-8q}
        \addplot+[mark=none, each nth point=20] table [x=steps, y=metropolis_4q, col sep=comma] {QCNCImages/data_grad_free.csv};
        \addlegendentry{Metro-4q}
        \addplot+[mark=none, each nth point=20] table [x=steps, y=metropolis_8q, col sep=comma] {QCNCImages/data_grad_free.csv};
        \addlegendentry{Metro-8q}

        \end{groupplot}
    \end{tikzpicture}
    \caption{Optimizer set comparison for circuit 3 in the predefined architectures for the bimodal GMM distribution. Solid lines: 4-qubits, dashed: 8-qubits. The size of the cost function landscape affects the performance of the optimizers, as the order drops significantly when scaled from 4 to 8-qubits. The steep descent of the better performing optimizers is due to the convergence protocols once a certain threshold is achieved.}
    \label{fig:optimizer_scaling_small}
\end{figure}


\begin{figure*}[htbp]
    \centering
    \begin{subfigure}[t]{0.3\textwidth}
        \centering
        \includegraphics[width=\linewidth]{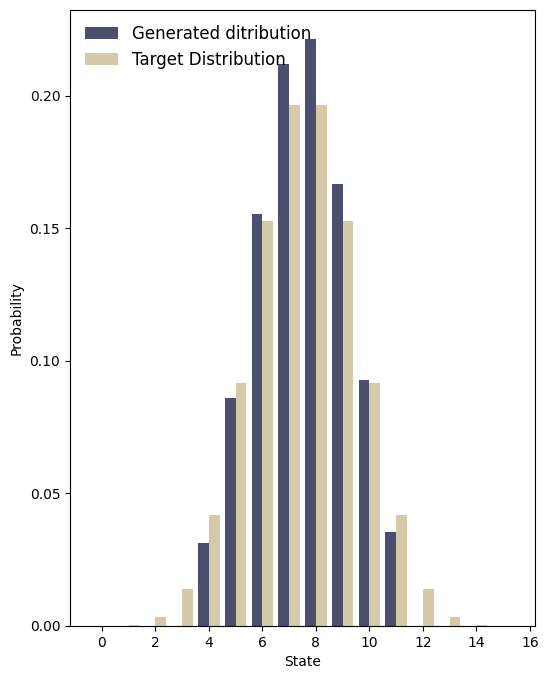}
        \caption{Normal Distribution using Predefined circuit (High expressibility)}
    \end{subfigure}\hfill
    \begin{subfigure}[t]{0.3\textwidth}
        \centering
        \includegraphics[width=\linewidth]{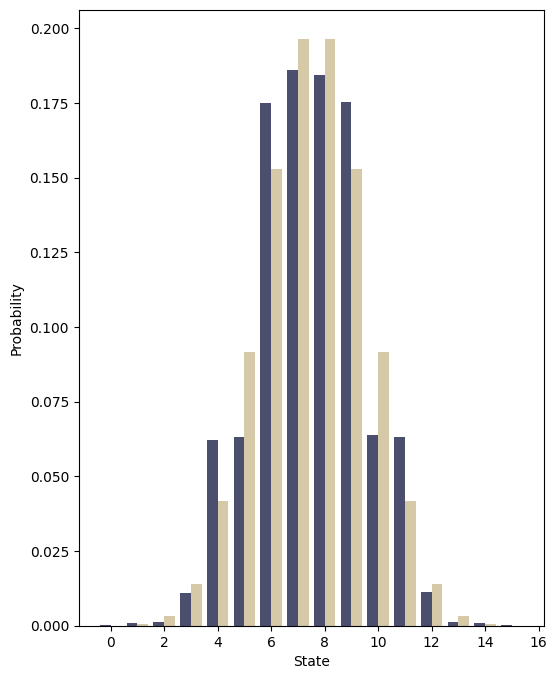}
        \caption{Normal Distribution using Enhanced Circuit 2}
    \end{subfigure}\hfill
    \begin{subfigure}[t]{0.3\textwidth}
        \centering
        \includegraphics[width=\linewidth]{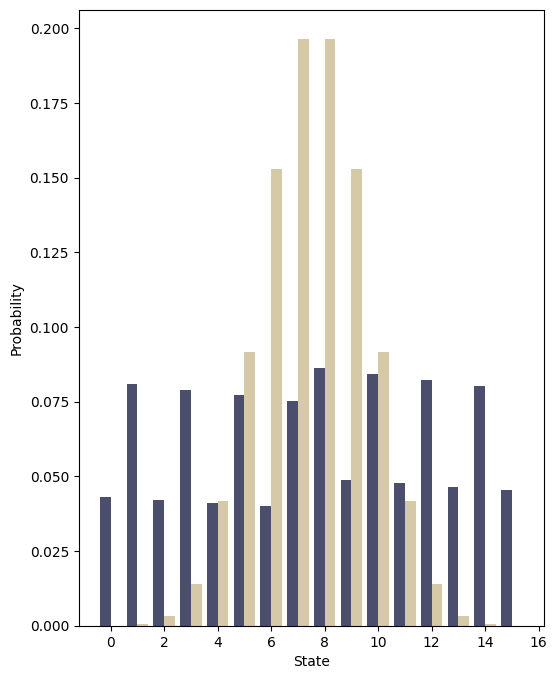}
        \caption{Normal Distribution using Random Circuit 2}
    \end{subfigure}

    \vspace{0.8em}

    \begin{subfigure}[t]{0.3\textwidth}
        \centering
        \includegraphics[width=\linewidth]{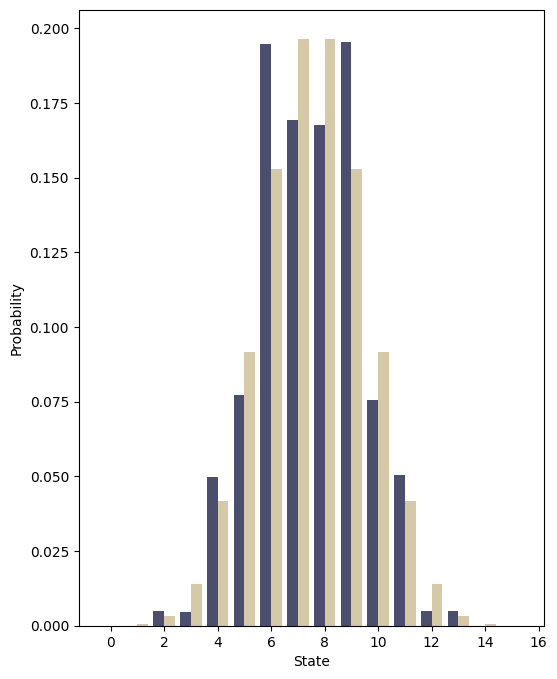}
        \caption{Normal Distribution using Enhanced Circuit 1}
    \end{subfigure}\hspace{0.05\textwidth}
    \begin{subfigure}[t]{0.3\textwidth}
        \centering
        \includegraphics[width=\linewidth]{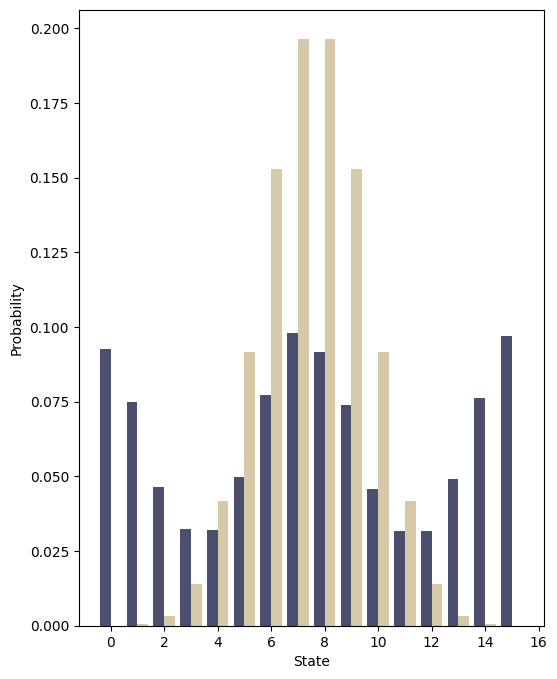}
        \caption{Normal Distribution using Random Circuit 1}
    \end{subfigure}
    \begin{subfigure}[t]{0.3\textwidth}
        \centering
        \includegraphics[width=\linewidth]{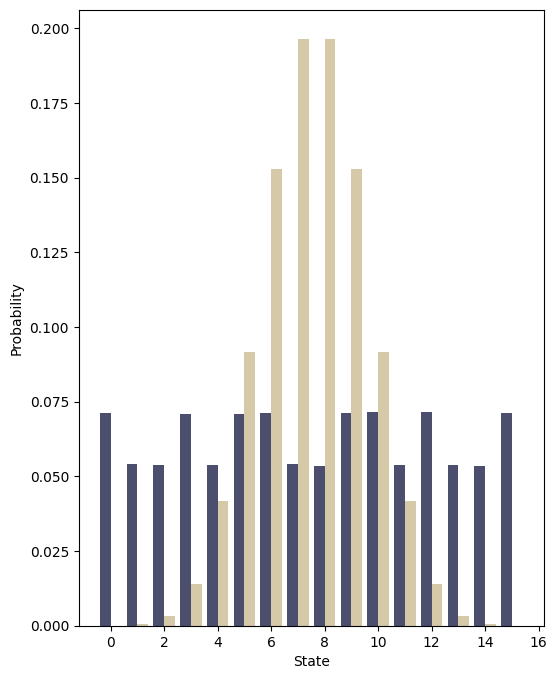}
        \caption{Normal Distribution using Predefined circuit (Low expressibility)}
    \end{subfigure}
    \caption{Target and best-produced distributions by 4 qubit circuits(best seed chosen by minimum JS Divergence)}
    \label{fig:best-dists}
\end{figure*}

\begin{figure}[ht]
    \centering
    \begin{tikzpicture}
        \begin{axis}[
            width=8cm,   
            height=8cm,  
            xmode=log, 
            ymode=log,
            grid=major,
            grid style={dotted, gray!50},
            xlabel={\textbf{Iterations}},
            ylabel={\textbf{Cost (JSD)}}, 
            tick label style={font=\scriptsize},
            label style={font=\footnotesize\bfseries},
            title style={font=\footnotesize\bfseries, yshift=-1ex},
            title={\textbf{Cost Function Comparison (Adam)}},
            xmin=1, xmax=1000, 
            ymin=1e-6, ymax=1e0, 
            legend style={
                at={(0.02,0.02)}, 
                anchor=south west,
                font=\large,
                draw=none, 
                fill=white, 
                fill opacity=0.8,
                cells={anchor=west}
            }
        ]

        \addplot[cBlue, thick] table [x=steps, y=jsd_4q, col sep=comma] {QCNCImages/data_metric_comparison.csv};
        \addlegendentry{JSD (4q)}
        \addplot[cBlue, thick, dashed] table [x=steps, y=jsd_8q, col sep=comma] {QCNCImages/data_metric_comparison.csv};
        \addlegendentry{JSD (8q)}

        \addplot[cOrange, thick] table [x=steps, y=kl_4q, col sep=comma] {QCNCImages/data_metric_comparison.csv};
        \addlegendentry{KL (4q)}
        \addplot[cOrange, thick, dashed] table [x=steps, y=kl_8q, col sep=comma] {QCNCImages/data_metric_comparison.csv};
        \addlegendentry{KL (8q)}

        \addplot[cRed, thick] table [x=steps, y=hellinger_4q, col sep=comma] {QCNCImages/data_metric_comparison.csv};
        \addlegendentry{Hell. (4q)}
        \addplot[cRed, thick, dashed] table [x=steps, y=hellinger_8q, col sep=comma] {QCNCImages/data_metric_comparison.csv};
        \addlegendentry{Hell. (8q)}

        \end{axis}
    \end{tikzpicture}
    \caption{Metric comparison from training Circuit 14 using adam optimizer for the GMM target distribution. Solid lines: 4-qubits, Dashed: 8-qubits.}
    \label{fig:metric_comparison}
\end{figure}

\begin{table}[ht]
    \centering
    \caption{Performance benchmarking of the better performing configuration (CMA-ES + JSD) across target distributions.}
    \label{tab:best_combo_performance}
    \vspace{-0.1cm}
    \renewcommand{\arraystretch}{1.2}
    \resizebox{\columnwidth}{!}{%
    \begin{tabular}{|c l c c c|}
        \hline
        \boldmath$n$ & \textbf{Target Distribution} & \textbf{Mean Cost} & \textbf{Best Cost} & \textbf{Std. Dev.} \\
        \hline
        
        \multirow{4}{*}{4} & Arbitrary & $0.0574$ & $3.50 \times 10^{-14}$ & $0.0926$ \\
 & Gmm Bimodal & $0.0733$ & $4.10 \times 10^{-14}$ & $0.1232$ \\
 & Normal & $0.0919$ & $7.83 \times 10^{-14}$ & $0.1589$ \\
 & Uniform & $0.0405$ & $1.35 \times 10^{-16}$ & $0.0870$ \\
 \cline{2-5}
\multirow{4}{*}{8} & Arbitrary & $0.0515$ & $0.0033$ & $0.0459$ \\
 & Gmm Bimodal & $0.0317$ & $9.50 \times 10^{-6}$ & $0.0645$ \\
 & Normal & $0.0482$ & $3.11 \times 10^{-6}$ & $0.0826$ \\
 & Uniform & $0.0103$ & $5.63 \times 10^{-14}$ & $0.0471$ \\
        
        \hline
    \end{tabular}%
    }
    
    \raggedright
    \vspace{0.8ex}
    \footnotesize{\textit{} $n$ denotes Number of Qubits. Mean Cost is evaluated over all architectures from our results. Best Cost is the convergence value of the best performing circuit. Standard Deviation represents the variance in the performance of the architectures.}
\end{table}

\subsection{Noisy simulation}
\label{sec:noise-model-short}

We model gate-level errors by inserting three channels after gates: (i) depolarizing, (ii) amplitude-damping and (iii) phase-damping \cite{nielsen_chuang}. Readout error is modelled as independent per-qubit bit flips applied to sampled bitstrings.

The noisy results in Table~\ref{tab:avg-js} and Fig.~\ref{fig:best-dists} show that Enhanced Circuit~2 outperforms the predefined high-expressibility circuit, while Enhanced Circuit~1 performs comparably. This can be attributed to the fact that the enhanced circuits were explicitly designed to achieve high expressibility under resource constraints. Thus, they retain sufficient expressive power while having lower circuit depth, making them less susceptible to noise. This is also reflected in their shorter runtimes (approximately 50 seconds) compared to the predefined circuit (approximately 100 seconds). In contrast, random circuits, although having similar depths and therefore similar noise effects, lack sufficient expressibility, failing to converge to the target distribution. These results highlight the importance of balancing expressibility and circuit depth for robust performance on noisy intermediate-scale quantum hardware.

\begin{table}[ht]
\centering
\caption{Average final cost and standard deviation by set and ID}
\label{tab:avg-js}
\begin{tabular}{lcrr}
\toprule
\textbf{Set} & \textbf{ID} & \textbf{Average final cost} & \textbf{Standard deviation} \\
\midrule
\multirow{2}{*}{Predefined} & 1 & 0.013624 & $2.01\times 10^{-5}$ \\
                           & 2 & 0.192814   & $4.03\times 10^{-7}$ \\
\cmidrule(lr){1-4}
\multirow{2}{*}{Random}     & 1 & 0.181456      & $3.70\times 10^{-4}$ \\
                           & 2 & 0.186998      & $2.31\times 10^{-4}$ \\
\cmidrule(lr){1-4}
\multirow{2}{*}{Enhanced}   & 1 & 0.013533      & $1.49\times 10^{-3}$ \\
                           & 2 & 0.008099      & $4.99\times 10^{-5}$ \\
\bottomrule
\end{tabular}
\end{table}

\newpage
\section{Conclusion}
\label{sec:concl}

In this work, we study how different variational circuit architecture perform for probability distribution generation, with a focus on role of expressibility, entanglement and quantum resource usage. We showed predefined structured circuits achieve high accuracy, their resource requirements grows rapidly with system size, limiting their scalability. We can achieve similar performance with much less resources by effectively utilizing the to enhance expressibility. Expressibility enhanced circuits achieve comparable fidelity using significantly fewer gates and shallower circuits making them more suitable for near-term quantum hardware. 

A key insight from our study is that high expressibility alone is not sufficient, the best performance is obtained by pairing high expressibility with sufficient entanglement capability. This provides a practical way to explain training behavior and guide ansatz selection. These results offer a concrete guidance for resource aware design of variational algorithms for sampling and quantum machine learning tasks on noisy devices. 

\newpage
\newpage
\bibliographystyle{IEEEtran}

\bibliography{newbib}

\end{document}